\documentclass[12pt,a4paper]{article}
\usepackage[hmargin=1.15in,vmargin=1in]{geometry}
\usepackage{amsmath,amssymb,amsfonts}
\usepackage{url}
\usepackage{graphicx,tikz,pgf,epsf}
\usepackage[ruled, vlined]{algorithm2e}
\begin{document}
\newcommand{\pf}{{\bf Proof: }}
\newcommand{\qed}{\hfill \rule{2mm}{2mm}}
\newtheorem{theorem}{Theorem}
\newtheorem{lemma}{Lemma}
\newtheorem{proposition}{Proposition}


\title{Designing Parity Preserving Reversible Circuits}
\author{Goutam Paul\thanks{The work of this author was done in part during his visit at RWTH Aachen, Germany as an Alexander von Humboldt Fellow.}\\
Department of Computer Science and Engineering,\\
Jadavpur University, Kolkata 700 032, India.\\
Email: goutam.paul@ieee.org
\and
Anupam Chattopadhyay\\
Institute for Communication Technologies and Embedded Systems,\\
RWTH Aachen University, Aachen 52074, Germany.\\
Email: anupam.chattopadhyay@ice.rwth-aachen.de
\and Chander Chandak\\
Electrical, Electronics and Communications Engineering,\\
Indian Institute of Technology Kharagpur, India.\\
Email: chandar.chandak@gmail.com
}
\date{}
\maketitle
\begin{abstract}
Making a reversible circuit fault-tolerant is much more difficult than 
classical circuit and there have been only a few works in the area of 
parity-preserving reversible logic design. Moreover, all of these designs are 
ad hoc, based on some pre-defined parity preserving reversible gates as 
building blocks. In this paper, we for the first time propose a novel and 
systematic approach towards parity preserving reversible circuits design. 
We provide some related theoretical results and give two algorithms, one from 
reversible specification to parity preserving reversible specification and 
another from irreversible specification to parity preserving reversible 
specification. We also evaluate the effectiveness of our approach by
extensive experimental results.
\end{abstract}


\section{Introduction and Motivation}
\begin{sloppypar}
It is known that erasure of a single bit of information dissipates heat 
equivalent to $K_BT\ln 2$~\cite{landauer61}, where 
$K_B = 1.38\times10^{-23}$ J/K is Boltzmann constant and $T$ is the room 
temperature in Kelvin. This heat dissipation is in conformity with the
laws of thermodynamics applied to any irreversible process.
Though classical logic is not reversible, it is
possible to represent classical Boolean functions using reversible
computation~\cite{bennett73}. On the other hand, any quantum computation is 
based on unitary evolution of quantum mechanical systems and is inherently 
reversible. However, with increasing demand on low power design, reversible 
logic finds application not only in quantum circuits, but also in
classical applications involving nanotechnology, optical circuits,
encoding/decoding etc.
\end{sloppypar}

\begin{sloppypar}
Any physical device performing classical or quantum computation is
subject to error due to noise in the environment or imperfections in the
device. {\em Fault tolerant computing} can mitigate this. One of the important
approaches towards fault tolerant computing is by using redundant
parity bits. For classical circuits, {\em bit flip} is the most common type of 
error. For quantum circuits, in addition to bit flip, there might be {\em phase 
flip} as well. In this paper, we consider bit flip errors only.
\end{sloppypar}

\begin{sloppypar}
Most common method for detecting bit-flip errors in storage or transmission is
by means of parity checking. Classically, most arithmetic and other processing 
functions do not preserve
the parity. One has to use redundant circuitry to compute and check
the parity. Making a reversible circuit fault-tolerant is
much more difficult than classical circuit, since reversible logic allows
no feedback or fan-out. In~\cite{parhami06}, the notion of 
{\em parity preserving reversible circuits} was introduced. The idea is to
design the reversible circuit in such a way that the parity between the input
and the output bits are automatically conserved in absence of any error.
\end{sloppypar}

\begin{sloppypar}
After~\cite{parhami06}, there has been a series of sporadic works in this area, 
such as designing adders~\cite{islam09}, divider~\cite{dastan11}, 
multiplier~\cite{qi12}, multiplexer~\cite{saligram13a}, ALU~\cite{saligram13b}
etc. However, all 
of these designs are ad hoc, based on some pre-defined parity preserving
reversible gates as building blocks. To the best of our knowledge, in this
paper, we for the first time propose a novel and systematic approach towards
parity preserving reversible circuits design. We provide some related
theoretical results and give two algorithms, one from reversible specification
to parity preserving reversible specification and another from 
irreversible specification to parity preserving reversible specification.
\end{sloppypar}

\section{Reversible Logic Synthesis}
\begin{sloppypar}
An $n$-variable Boolean function is \textit{reversible} if all its output patterns map uniquely to an input pattern and vice-versa. It can be expressed as an $n$-input, 
$n$-output bijection or alternatively, as a permutation function over the truth value set $\{0, 1, \ldots 2^{n-1}\}$. The problem of reversible logic synthesis is to map 
such a reversible Boolean function on a reversible logic gate library.
\end{sloppypar}

\begin{sloppypar}
The gates are characterized by their implementation cost in quantum technologies, which is dubbed as Quantum Cost (QC) \cite{miller_ismvl11}\cite{maslov_benchmark}. 
Reversible logic gates can also be represented as an unitary transformation, therefore serving as building blocks for quantum computers. Few prominent classical reversible 
logic gates are presented below.
\end{sloppypar}

\begin{itemize}
\item NOT gate: $f(A)$ = $\overline{A}$.
\item CNOT gate: $f(A)$ = $A$, $f(B)$ = $A \oplus B$.
\item CCNOT gate: Also known as Toffoli gate. $f(A)$ = $A$, $f(B)$ = $B$, $f(C)$ = $AB \oplus C$. This gate can be generalized with $Tof_n$ gate, where first $n-1$ 
variables are used as control lines. NOT and CNOT gates are denoted as $Tof_1$ and $Tof_2$ respectively.
\item Peres gate: A sequence of $Tof_3(a, b, c)$, $Tof_2(a, b)$ or its inverse is known as Peres gate.
\item Controlled Swap gate: Also known as Fredkin gate. $f(A)$ = A, $f(B)$ = $\overline{A}.B + A.C$, $f(C)$ = $\overline{A}.C + A.B$. This gate can be 
generalized with $Fred_n$ gate $(n > 1)$, where first $n-2$ variables are used as control lines.
\end{itemize}

\begin{sloppypar}
Multiple sets of reversible gates form an universal gate library for realizing classical Boolean functions such as, (i) NCT: NOT, CNOT, Toffoli. (ii) NCTSF: NOT, CNOT, 
Toffoli, SWAP, Fredkin. (iii) GT: $Tof_n$. (iv) GTGF: $Tof_n$ and $Fred_n$.
\end{sloppypar}

\begin{sloppypar}
Reversible logic synthesis begins from a given $n$-variable Boolean function, which can be irreversible. The first step is to convert it to a reversible Boolean function 
by adding distinguishing output bits, known as \textit{garbage outputs}. When additional input Boolean variables are needed for constructing the output function, those 
are referred as \textit{ancilla}.
\end{sloppypar}

\begin{sloppypar}
Reversible logic synthesis methods can be broadly classified in four categories as following. A different and more detailed classification is presented in a recent survey 
of reversible logic synthesis methods \cite{markov_survey}. \\
\textbf{Exact and Optimal methods:} These methods consider step-by-step 
exhaustive enumeration or formulating the logic synthesis as a SAT problem \cite{wille_sat} or reachability problem \cite{reachability_hung}. 
Optimal implementation up to only $4$-variable Boolean functions are known~\cite{golubitsky_4var}. \\
\textbf{Transformation-based method \cite{mmd}\cite{rgraph}:} 
These methods use a weighted graph representation for performing the transformations, while \cite{mmd} proceed row-wise in the Boolean truth-table.\\
\textbf{Methods based on decision diagrams \cite{wille_bdd,qmdd}:} In this
approach, each node of the decision diagram is converted to an equivalent reversible circuit 
structure. These methods reported excellent scaling for large Boolean functions, low QC at the cost of high number of garbage bits. \\
\textbf{ESOP-based methods:} For classical logic synthesis, the exclusive sum
of products (ESOP) formulation is studied well for specific target technologies~\cite{xorcism}. For reversible logic synthesis, the ESOP formulation~\cite{esop_gupta} maps directly 
to the basic reversible logic gates and has led to significant research interest. 
\end{sloppypar}

\begin{sloppypar}
Among the above methods, methods based on Decision Diagrams and ESOP-based methods can synthesize an Irreversible Boolean specification to reversible 
circuit by adding extra garbage lines. However, these methods do not guarantee the minimum garbage count. On the other hand, determination of minimum 
garbage count and their assignment is non-trivial, particularly for Boolean functions with large number of variables~\cite{wille_min_garbage}. To the best of 
our knowledge, no automatic reversible logic synthesis tool supports automatic derivation of parity-preserving Boolean specification from an irreversible/reversible 
Boolean specification. Our flow proposed in the paper can be complemented with any reversible logic synthesis flows, which work on reversible Boolean specifications.
\end{sloppypar}

\section{Our Results}
First we discuss how to convert a reversible Boolean specification (that does
not necessarily consider parity preservation) into parity-preserving reversible 
specification. Before proceeding, we count the number of $n$-variable parity 
preserving reversible Boolean functions in Theorem~\ref{count}.
\begin{theorem}
\label{count}
Total number of $n$-variable parity preserving reversible Boolean functions
is $\left(2^{n-1}!\right)^2$.
\end{theorem}
\pf In the truth table of an $n$-variable reversible Boolean function, 
there are $2^n$ input and output rows. Half of the $2^n$ input (or output) 
rows, i.e., total $2^{n-1}$ rows would have odd parity and the other half would 
have even parity. For the function to be parity-preserving, the odd-parity 
input rows must map to the odd-parity output rows. There are $2^{n-1}!$ such 
mappings.  Corresponding to each of these, the even-parity input rows must map 
to the even-parity output rows and there are again $2^{n-1}!$ such mappings.
Hence the result follows.  \qed

The method of constructing a parity-preserving reversible specification from
any reversible specification is described in the proof of Theorem~\ref{rev2rev}.
\begin{theorem}
\label{rev2rev}
Given any $n$-variable reversible Boolean specification, it can be converted
to a parity-preserving reversible Boolean specification with the introduction 
of at most one extra variable.
\end{theorem}
\pf If the function is already parity-preserving, we need not do anything.
If not, then in the output column of the truth table, we can just put a 0 in 
the parity-matching rows and a 1 in the parity-mismatching rows. On the input 
side, the extra variable can be set to the constant 0. Hence the result 
follows. \qed

\subsection{Direct Method of Converting Irreversible Specification to
Parity-preserving Reversible Specification}
\begin{sloppypar}
Next, we discuss the case when we are given an irreversible Boolean
specification. One simple approach can be a two-phase procedure: first,
to use some standard approaches \cite{wille_min_garbage} for converting the irreversible 
specification to a reversible specification, and next, use the result of
Theorem~\ref{rev2rev}. However, the first phase in this approach may 
incur unnecessary extra garbage bits. To avoid this problem, we provide
a direct method of converting a given irreversible specification to 
a parity-preserving reversible specification with theoretically bounded
number of extra bits. The method is as follows.
\end{sloppypar}

\begin{sloppypar}
Since the specification is irreversible, the output rows must contain
duplicate bit-strings. Suppose there are $n$ input variables and hence
$2^n$ rows in the truth table. Suppose there are $k < 2^n$ distinct
output bit-strings, with the counts $n_1, \ldots, n_k$, such that
$\sum_{i=1}^{k} n_i = 2^n$. For each $i = 1, \ldots, k$, out of $n_i$ rows 
with the same output bit-string, let $n_{i,p}$ be the number of rows where the 
input and the output parity is matching and so $n_i - n_{i,p}$ is the number of 
rows where the parity is not matching. To differentiate the matching rows
we need at least $\lceil \log_2 n_{i,p}\rceil$ extra bits. 
Similarly, to differentiate the mismatching rows, we need at least 
$\lceil \log_2 \left(n - n_{i,p}\right)\rceil$ extra bits. Hence, for
the rows corresponding to the bit-string category $i$, the number of
extra bits needed is one more than the maximum of these two numbers. The one
additional bit is required to match the parity.
Thus, the total
number of extra bits needed is given by the maximum of the above quantity 
over all $i$'s. Hence, with the above formulation, we have the following result.
\end{sloppypar}
\begin{theorem}
\label{irrev2rev}
The minimum number of extra bits needed to convert an irreversible 
specification to parity-preserving reversible specification is given by 
$$\displaystyle\max_{i=1}^{k} \max\{\lceil \log_2 n_{i,p}\rceil+1, \lceil \log_2 \left(n - n_{i,p}\right)\rceil+1\}.$$
\end{theorem}

\subsection{Algorithm and its Complexity Analysis}
In Algorithm~\ref{GenParityRev}, we present the procedure for converting an 
irreversible specification to parity-preserving reversible specification.
Suppose $x_1, \ldots, x_k$ are $k$ integers 
$\in \{0, \ldots, 2^n-1\}$ denoting the decimal equivalent of
distinct output bitstrings. Note that according to our notation,
$x_i$ appears $n_i$ times. We will keep two arrays $match$ and $mismatch$
as follows. In the algorithm, $match[x_i]$ will contain $n_{i,m}$ and
$mismatch[x_i]$ will contain $n-n_{i,m}$.
The array $count[i]$, for $0, \ldots, 2^n-1$, is filled from top to 
bottom order, corresponding to each output row as follows: $count[i]$
contains how many times the $i$-th output row has
appeared so far starting from the top row. The sign of $count[i]$ is positive, 
if the parity is preserved, else it is negative.

\begin{algorithm}[htb]
\KwIn{$n$, An integer array $out[0 \ldots 2^{n}-1]$, containing the decimal 
equivalent of the output rows of an $n$-variable Boolean function.}
\KwOut{Parity preserving reversible specification.}
\nl $max = 0$\;
\lnl{start1} \For{$i = 0$ to $2^n - 1$} {
\nl 	$match[i] = 0$, $mismatch[i] = 0$, $count[i] = 0$\;
} 
\nl \For{$row \leftarrow 0$ \KwTo $2^n-1$} {
\nl If parity matches, increment $match[out[row]]$ by 1\;
\nl Otherwise, decrement $mismatch[out[row]]$ by 1\;
\nl  \If{$max < match[out[row]]$}{
\nl  $max = match[out[row]]$, $count[row] = match[out[row]]$\;
  }
\nl\If{$max < mismatch[out[row]]$}{
\lnl{end1} $max = match[out[row]]$, $count[row] = - match[out[row]]$\;
  }
}

\nl $g = \log _2 max + 1$\;
\nl Add $g$ columns to the Boolean output specification\;
\lnl{start2} \For{$row \leftarrow 0$ \KwTo $2^n-1$} {
\nl  $k = abs(count[row])$\;
\nl  Append binary value of $k$ in the $g-1$ bits\;
\lnl{end2} Use the last bit to match parity\;
}
\caption{Irreversible to Parity Preserving Reversible Specification}
\label{GenParityRev}
\end{algorithm}

Now we present the complexity of our algorithm in Theorem~\ref{comp}.
\begin{theorem}
\label{comp}
For an $n$-input $m$-output Boolean specification, the running time of 
Algorithm~\ref{GenParityRev} is $O((n+m)2^n)$.
\end{theorem}
\pf The maximum number of input or output rows in the Boolean specification is 
$2^n$. Let there be $k < 2^n$ distinct output bit-strings with the counts 
$n_1, \ldots, n_k$, such that $\sum_{i=1}^{k} n_i = 2^n$. 
For each row we have to compute the number of $1$'s in the input and output 
bit-strings for computing the parity.
The algorithmic complexity for this traversal is $O((n+m)2^n)$, which
accounts for Steps~\ref{start1} to~\ref{end1}.
After this computation, we have one more iteration over the output rows
through Step~\ref{start2} to~\ref{end2}, the running time of which 
is dominated by $O((n+m)2^n)$. Hence the result follows. \qed

\section{Experimental Results}
\begin{sloppypar}
The proposed algorithm has been implemented and tested on several
benchmark circuits, using C++ on an
Intel(R) Core(TM) i5-3570 CPU (Quad-core) with 3.40GHz clock and 6 MB cache,
having Linux version 2.6.32-358.6.2.el6.x86\_64 as the OS, and 
gcc version 4.4.7 as the compiler.
First, we compared 
our automatically generated parity-preserving reversible circuits with manually created parity-preserving reversible circuits reported by others. 
Our comparison metric is the number of additional garbage lines required for preserving parity.
\end{sloppypar}

\subsection{Comparison with State-of-the-art}
\begin{sloppypar}
After following the proposed algorithm the irreversible Boolean specification is transformed to 
a reversible one (Table \ref{table:half_adder}, Table \ref{table:full_adder}) with the required number of constant input and garbage lines. 
The ancilla inputs and garbage outputs are referred as $A_i$  and $G_i$ respectively. The
reversible specification thus obtained can be used to implement the reversible circuit using the 
well-known reversible logic synthesis methods for garbage-free synthesis~\cite{mmd}. 
\end{sloppypar}

{\small
\begin{table}[htbp]
  \caption{Half adder Boolean Specification}
  \label{table:half_adder}
  \centering
  \begin{tabular}{c|c|ccc|ccc}
    \hline
    \multicolumn{2}{c|}{Irreversible Specification} & \multicolumn{6}{|c}{Reversible Specification} \\ \hline
    $Input$ & $Output$ & $Input$ & $A_1$ & $A_2$ & $Output$ & $G_1$ & $G_2$ \\    \hline
    00 & 00 & 00 & 0 & 0 & 00 & 0 & 0 \\
    01 & 10 & 01 & 0 & 0 & 10 & 0 & 0 \\
    10 & 10 & 10 & 0 & 0 & 10 & 1 & 1 \\
    11 & 01 & 11 & 0 & 0 & 01 & 0 & 1 \\    
    \hline
  \end{tabular}
\end{table}
}

{\small
\begin{table}[btbp]
  \caption{Full Adder Boolean Specification}
  \label{table:full_adder}
  \centering
  \begin{tabular}{c|c|ccc|cccc}
    \hline
    \multicolumn{2}{c|}{Irreversible Specification} & \multicolumn{6}{|c}{Reversible Specification} \\ \hline
    $Input$ & $Output$ & $Input$ & $A_1$ & $A_2$ & $Output$ & $G_1$ & $G_2$ & $G_3$ \\     \hline
    000 & 00 & 000 & 0 & 0 & 00 & 0 & 0 & 0 \\
    001 & 10 & 001 & 0 & 0 & 10 & 0 & 0 & 0 \\
    010 & 10 & 010 & 0 & 0 & 10 & 0 & 1 & 1 \\
    011 & 01 & 011 & 0 & 0 & 01 & 0 & 0 & 1 \\
    100 & 10 & 100 & 0 & 0 & 10 & 1 & 0 & 1 \\
    101 & 01 & 101 & 0 & 0 & 01 & 0 & 1 & 1 \\
    110 & 01 & 110 & 0 & 0 & 01 & 1 & 0 & 1 \\
    111 & 11 & 111 & 0 & 0 & 11 & 0 & 0 & 1 \\
    \hline
  \end{tabular}
\end{table}
}

\begin{sloppypar}
In terms of the ancilla and garbage count, we obtain exactly the same number for both 
the half-adder and full-adder circuits as obtained manually in~\cite{comparison_parity,azad_02}.
\end{sloppypar}


\subsection{Testing for Boolean functions with Large Variable Count}

\begin{sloppypar}
Apart from this we had tried the algorithm for several Boolean functions with large number of variables, 
for which obtaining a parity-preserving Boolean specification manually would be hard. These are presented 
in Table \ref{table:time_comparison}. In the table, the tar\_* functions are from Tarannikov's paper~\cite{Tarannikov}. 
From \cite[Equation 2]{Tarannikov}, we use the parameter $c$ as {001} to construct an $8$-variable, $2$-resilient function 
then we get tar82\_2\_001.pla. Similarly tar93\_110.pla and tar93\_101.pla are 9 variable 3-resilient functions 
with the $c$ vector as {110} and {101} respectively. The functions like $rdNK$ is presented in several benchmarks on 
reversible logic synthesis~\cite{maslov_benchmark}. The input weight function $rdNK$ has $N$ inputs and $K= {\lfloor logN \rfloor + 1 }$ 
outputs. Its output is the binary encoding of the number of ones in its input. The other functions are obtained from RevKit benchmark~\cite{revkit}.
\end{sloppypar}

{\scriptsize
\begin{table}[htbp]
  \caption{Summary of results for exemplary Boolean functions with large no. of variables}
  \label{table:time_comparison}
  \centering
  \begin{tabular}{c|c|c|c|c|c}
    \hline
    Function  & Input & Output & Garbage & Ancilla & Runtime (ms) \\
    \hline
    tar82\_2\_001.pla & 8 & 1 & 8 & 1  & 0.657 \\
    tar93\_110.pla & 9 & 1 & 8 & 0 & 1.888 \\
    tar93\_101.pla & 9 & 1 & 8 & 0  & 1.631 \\
    rd53 & 5 & 3 & 5 & 3 & 0.18 \\
    rd73 & 7 & 3 & 7 & 3 & 0.35 \\
    rd84 & 8 & 4 & 8 & 4& 0.64 \\
    rd20\_5 & 20 & 5 & 19 & 4  & 34.698 \\
    rd10\_4 & 10 & 4 & 9 & 3  & 23.175 \\
    0410184\_85.pla & 14 & 14 & 1 & 1  & 14.172 \\
    cycle10\_2\_61.pla & 12 & 12 & 1 & 1  & 3.394 \\
    ham15\_30.pla & 15 & 15 & 1 & 1  & 30.152 \\
    ham7\_29.pla & 7 & 7 & 1 & 1  & 0.198 \\
    ham8\_64.pla & 8 & 8 & 1 & 1  & 0.314 \\
    life\_175.pla & 9 & 1 & 9 & 1  & 0.448 \\
    squar5.pla & 5 & 8 & 1 & 4  & 6.765 \\
    urf4\_89.pla & 11 & 11 & 1 & 1  & 1.76 \\
    urf6.pla & 15 & 15 & 1 & 1  & 29.208 \\
    plus63mod8192.pla & 13 & 13 & 1 & 1  & 6.757 \\
    \hline
  \end{tabular}
\end{table}
}

\begin{sloppypar}
Our proposed algorithm can be used on any irreversible specification unlike the
methods described in ~\cite{comparison_parity} and ~\cite{islam09} where a new specific 
gate is introduced to realize one particular circuit. These gates may not be useful to 
realize other circuits. Our method is fully automated and general.
\end{sloppypar}

\end{document}